\documentclass[aps,prl,showpacs,showkeys,twocolumn,amssymb,groupedaddress]{revtex4}

\usepackage{graphicx}% Include figure files
\usepackage{dcolumn}% Align table columns on decimal point
\usepackage{bm}% bold math

\begin{document}

\title{Black hole mass decreasing due to phantom energy accretion}

\author{E. Babichev}
 \email{babichev@inr.npd.ac.ru}
\author{V. Dokuchaev}
 \email{dokuchaev@inr.npd.ac.ru}
\author{Yu. Eroshenko}
 \email{erosh@inr.npd.ac.ru}

\affiliation{Institute For Nuclear Research of the Russian Academy of
Sciences \\
60th October Anniversary Prospect 7a, 117312 Moscow, Russia}

\date{\today}

\begin{abstract}
Solution for a stationary spherically symmetric accretion of the
relativistic perfect fluid with an equation of state $p(\rho)$
onto the Schwarzschild black hole is presented. This solution is a
generalization of Michel solution and applicable to the problem of
dark energy accretion. It is shown that accretion of phantom
energy is accompanied with the gradual decrease of the black hole
mass. Masses of all black holes tend to zero in the phantom energy
universe approaching to the Big Rip.
\end{abstract}

\pacs{04.70.Bw, 04.70.Dy, 95.35.+d, 98.80.Cq}

\keywords{black holes, cosmology, dark energy}
%Use showkeys class option if keyword display desired

\maketitle

Observations of distant supernovae and cosmic microwave background
anisotropy indicate in favor of the accelerating expansion of the
Universe \cite{acceler}.  In the framework of a general relativity
it means that a considerable part of the Universe consists of the
dark energy:  the component with a positive energy density
$\rho>0$ and with a negative pressure $p<-(1/3)\rho$.  This dark
energy may be in the form of a vacuum energy (cosmological
constant $\Lambda$) with $p=-\rho$, or a dynamically evolving
scalar field with a negative pressure (quintessence \cite{CaDaSt}
or $k$-essence \cite{ArMuSt}). One of the peculiar feature of the
cosmological dark energy is a possibility of the Big Rip
\cite{Caldw}:  the infinite expansion of the universe during a
finite time.  The Big Rip scenario is realized if a dark energy is
in the form of the phantom energy with $\rho+p<0$.  In this case
the cosmological phantom energy density grows at large times and
disrupts finally all bounded objects up to subnuclear scale.

What would be the fate of black holes in the universe filled with
the phantom energy?  Below we find the solution for a stationary
accretion of the relativistic perfect fluid with an arbitrary
equation of state $p(\rho)$ onto the Schwarzschild black hole.
Using this solution we show that the black hole mass diminishes by
accretion of the phantom energy.  Masses of all black holes
gradually tend to zero in the phantom energy universe approaching
to the Big Rip. The diminishing of a black hole mass is caused by
the violation of the energy domination condition $\rho+p\geq0$
which is a principal assumption of the classical black hole
`non-diminishing' theorems \cite{hawkell}. The another consequence
of the existence of a phantom energy is a possibility of
traversable wormholes \cite{wormh}.

We model the dark energy in the black hole gravitational field by
the test perfect fluid with a negative pressure, an arbitrary
equation of state $p(\rho)$ and energy-momentum tensor
$T_{\mu\nu}=(\rho+p)u_\mu u_\nu - pg_{\mu\nu}$, where $\rho$ and
$p$ are the dark energy density and pressure correspondingly, and
$u^\mu=dx^\mu/ds$ is a fluid four-velocity with
$u^{\mu}u_{\mu}=1$. The known analytic relativistic accretion
solution onto the Schwarzschild black hole by Michel \cite{Mich}
(see also \cite{Beg}) is not applied for a general case of a dark
energy accretion. We adjust the Michel solution for the problem of
dark energy accretion by elimination from the equations the
particle number density.

The integration of the time component of the energy-momentum
conservation law $T^{\mu\nu}_{\;\;\; ;\nu}=0$ gives the first
integral of motion for the stationary spherically symmetric
accretion (the relativistic Bernoulli or energy conservation
equation):
\begin{equation}
 \label{eq1}
  (\rho+p)\left(1-\frac{2}{x}+u^2\right)^{1/2}x^2 u =C_1,
\end{equation}
where $x=r/M$, $u=dr/ds$, $r$ is a radial Schwarzschild
coordinate, $M$ is a black hole mass and $C_1$ is a constant
determined below.  To obtain an another integral of motion we use
the projection of the energy-momentum conservation law on the
four-velocity $u_{\mu}T^{\mu\nu}_{\quad ;\nu}=0$, which for a
perfect fluid is
\begin{equation}
 \label{eq2}
  u^{\mu}\rho_{,\mu}+(\rho+p)u^{\mu}_{\; ;\mu}=0.
\end{equation}
The integration of (\ref{eq2}) gives the second integral of motion
(the energy flux equation):
\begin{equation}
 \label{flux}
 ux^2\exp\left[\;\,\int\limits_{\rho_{\infty}}^{\rho}\!\!
 \frac{d\rho'}{\rho'+p(\rho')}\right]=-A,
\end{equation}
where $u<0$ in the case of inflow motion and a dimensionless
constant $A>0$. Note that the second integral of motion
(\ref{flux}) is obtained without use of the particle conservation
law. From (\ref{eq1}) and (\ref{flux}) one can easily obtain:
\begin{equation}
 \label{energy}
 (\rho+p)\left(1-\frac{2}{x}+u^2\right)^{1/2}
 \exp\left[-\!\!\int\limits_{\rho_{\infty}}^{\rho}\!\!
 \frac{d\rho'}{\rho'+p(\rho')}\right]=C_2,
\end{equation}
where $C_2=-C_1/A=\rho_\infty+p(\rho_\infty)$. From (\ref{flux})
and (\ref{energy}) one can find the relations for the fluid
velocity $u_{\rm H}=u(2M)$ and density $\rho_{\rm H}=\rho(2M)$ at
the black hole horizon $r=2M$:
$$
 \frac{A}{4}\,
 \frac{\rho_{\rm H}+p(\rho_{\rm H})}{\rho_\infty+p(\rho_\infty)}
 =\frac{A^2}{16u_H^2}=\exp\left[2\,\int\limits_{\rho_{\infty}}^{\rho_H}\!\!
 \frac{d\rho'}{\rho'+p(\rho')}\right].
$$
The constant $A$ which determines the accretion flux is calculated
by fixing parameters at a critical point. This provides us with
the continuity of the solution from the infinity to a horizon.
Following Michel \cite{Mich} we fix parameters of critical point
$x=x_*$:
\begin{equation}
 \label{cpoint}
  u_*^2=\frac{1}{2x_*},\quad c_s^2(\rho_*)=\frac{u_*^2}{1-3u_*^2},
\end{equation}
where $c_s^2(\rho)=\partial p/\partial\rho$ is the square of sound
speed. Using (\ref{cpoint}) and (\ref{energy}) one can find the
following relation:
$$
% \label{rho_c}
 \frac{\rho_*+p(\rho_*)}{\rho_\infty+p(\rho_\infty)}
  \left[1+3c_s^2(\rho_*)\right]^{-1/2}
 =\exp\left[\;\,\int\limits_{\rho_{\infty}}^{\rho_*}\!\!
 \frac{d\rho'}{\rho'+p(\rho')}\right],
$$
which determines the density at a critical point
$\rho_*=\rho(x_*)$. Then for a given $\rho_*$, with the help of
(\ref{cpoint}), one can find $x_*$ and $u_*$. Constant $A$ is
fixed by substituting of the calculated values in (\ref{flux}).
Note that there is no critical point outside the black hole
horizon ($x_*>1$) for $c_s^2<0$ or $c_s^2>1$.  This means that for
unstable perfect fluid with $c_s^2<0$ or $c_s^2>1$ a dark energy
flux onto the black hole depends on the initial conditions. This
result has a simple physical interpretation: the accreting fluid
has the critical point if its velocity increases from subsonic to
trans-sonic values.  In a fluid with a negative $c_s^2$ or with
$c_s^2>1$ the fluid velocity never crosses such a point. It should
be stressed, however, that fluids with $c_s^2<0$ are
hydrodynamically unstable (see discussion in
\cite{FabMar97,Sean03}).

Equations (\ref{flux}) and (\ref{energy}) along with the equation
of state $p=p(\rho)$ describe the requested accretion flow onto
the black hole. These equations are valid for perfect fluid with
an arbitrary equation of state $p=p(\rho)$, in particular, for a
gas with zero-rest-mass particles (thermal radiation), for a gas
with nonzero-rest-mass particles and for a dark energy.  For a
nonzero-rest-mass gas the couple of equations (\ref{flux}) and
(\ref{energy}) is reduced to similar ones found by Michel
\cite{Mich}.

The black hole mass changes at a rate $\dot M=-4\pi r^2T_0^{\;r}$
due to the fluid accretion. With the help of (\ref{flux}) and
(\ref{energy}) this can be expressed as
\begin{equation}
 \label{evol}
 \dot{M}=4\pi A M^2 [\rho_{\infty}+p(\rho_{\infty})].
\end{equation}
From this equation it is clear that the accretion of a phantom
energy with $\rho_{\infty}+p(\rho_{\infty})<0$ is always
accompanied with the diminishing of the black hole mass.  This
result is valid for any equation of state $p=p(\rho)$ with
$p+\rho<0$. If we neglect the cosmological evolution of
$\rho_\infty$ then from (\ref{evol}) we obtain:
\begin{equation}
 \label{m}
 M=M_i\left(1-\frac{t}{\tau}\right)^{-1},
\end{equation}
where $M_i$ is an initial mass of the black hole and a
characteristic evolution time $\tau=\{4\pi A M_i
[\rho_\infty+p(\rho_\infty)]\}^{-1}$.

As a particular fully solvable example we consider the perfect
fluid with a linear equation of state:
\begin{equation}
 \label{p1}
 p=\alpha(\rho-\rho_0), \quad \alpha=const, \quad \rho_0=const,
\end{equation}
where $\alpha=c_s^2$ and it is supposed that $\rho>0$. Among
others this model includes radiation ($p=\rho/3$), ultra-hard
equation of state ($p=\rho$) and the simplest models of dark
energy ($\rho_0=0$, $\alpha<0$). The constant $\alpha$ is
connected with an often used parameter $w=p/\rho$ by the relation
$w=\alpha(\rho-\rho_0)/\rho$. The physically reasonable case
corresponds to $0<\alpha\leq 1$.

In the linear model (\ref{p1}) the radius of critical point
$x_*=(1+3\alpha)/2\alpha$, velocity at critical point
$u_*^2=\alpha/(1+3\alpha)$, velocity at the horizon $u_{\rm
H}=-(A/4)^{-\alpha/(1-\alpha)}$ and constant $A$ determining the
energy flux in (\ref{evol}) is \cite{bde}
\begin{equation}
\label{A1}
A=\frac{(1+3\alpha)^{(1+3\alpha)/2\alpha}}{4\alpha^{3/2}}.
\end{equation}
This relation is valid only for stable fluid with $0<\alpha\leq
1$. For unstable fluid with $\alpha<0$ the constant $A$ is
indeterminable by the condition of the solution continuity from
critical point consideration. We suppose that in the unstable case
the growth of instabilities in the accretion flow will cause the
asymptotic growth of the accretion velocity up to the limiting
speed of light at the horizon $u_{\rm H}\to-1$. This helps to fix
the value of constant $A$: $A=4$.

For some particular choices of parameter $\alpha$ the values
$\rho(x)$ and $u(x)$ can be calculated analytically. For example,
for $\alpha=1/3$ the fluid density is given by:
\begin{equation}
 \label{sol1}
 \rho=\frac{\rho_0}{4}+\left(\rho_{\infty}-\frac{\rho_0}{4}\right)
 \left[z+\frac{1}{3(1-2x^{-1})}\right]^2,
\end{equation}
where
$$
 z=\left\{ \begin{array}{ll}
 2{\sqrt{\frac{a}{3}}}\,\cos\left(\frac{2\,\pi }{3}
 -\frac{\beta}{3}\right),& 2\leq x\leq3,\\
 2{\sqrt{\frac{a}{3}}}\,\cos\left(\frac{\beta}{3}\right),& x>3,
\end{array} \right.
$$
$$
 \beta=\arccos\left[\frac{b}{2\,(a/3)^{3/2}}\right]
$$
and
$$
 a=\frac{1}{3{\left( 1 - \frac{2}{x}\right) }^2},\;
 b=\frac{2}{27{\left(1-\frac{2}{x}\right) }^3}- \frac{108}{\left(1-\frac{2}{x}\right) x^4}.
$$
The density distribution for another physically interesting case
$\alpha=1$ is given by:
\begin{equation}
 \label{sol2}
 \rho=\frac{\rho_0}{2}+\left(\rho_{\infty}-\frac{\rho_0}{2}\right)
 \left(1+\frac{2}{x}\right)\left(1+\frac{4}{x^2}\right).
\end{equation}
The corresponding radial fluid velocity $u=u(x)$ can be calculated
by substituting of (\ref{sol1}) or (\ref{sol2}) into (\ref{eq1}).
For $\rho_0=0$ the solutions (\ref{sol1}) and (\ref{sol2})
describe correspondingly a thermal radiation and a fluid with
ultra-hard equation of state. In the case of
$\rho_\infty<\alpha\rho_0/(1+\alpha)$ the solutions (\ref{sol1})
and (\ref{sol2}) describe the phantom energy  falling onto the
black hole. For example, a phantom energy flow with parameters
$\alpha=1$ and $\rho_0=(7/3)\rho_\infty$ results in a black hole
mass diminishing with the rate $\dot M=-(8\pi/3)(2
M)^2\rho_\infty$.

Now we turn to the problem of the black hole evolution in the
universe with the Big Rip when a scale factor $a(t)$ diverges at
finite time \cite{Caldw}. For simplicity we will take into account
only dark energy and will disregard all others forms of energy.
The Big Rip solution is realized for $\rho+p<0$ and $\alpha<-1$.
From the Friedman equations for the linear equation of state model
one can obtain: $|\rho+p|\propto a^{-3(1+\alpha)}$. Taking for
simplicity $\rho_0=0$ we find the evolution of the density of a
phantom energy in the universe:
\begin{equation}
\label{sol3}
\rho_\infty=\rho_{\infty,i}\left(1-\frac{t}{\tau}\right)^{-2},
\end{equation}
where
\begin{equation}
\label{tau} \tau^{-1}=-\frac{3(1+\alpha)}{2}\left(\frac{8\pi
G}{3}\rho_{\infty,i}\right)^{1/2}
\end{equation}
and $\rho_{\infty,i}$ is the initial density of the cosmological
phantom energy and the initial moment of time is chosen so that
the `doomsday' comes at time $\tau$. From (\ref{evol}) using
(\ref{sol3}) we find the black hole mass evolution:
\begin{equation}
 \label{mevol1}
 M=M_i\left(1+\frac{M_i}{\dot M_0 \;\tau}\;
 \frac{t}{\tau-t}\right)^{-1},
\end{equation}
where  $\dot M_0=(3/2)\,A^{-1}|1+\alpha|$ and $M_i$ is the initial
mass of the black hole. For $\alpha=-2$ and typical value of $A=4$
(corresponding to $u_{\rm H}=-1$) we have $\dot M_0=-3/8$. In the
limit $t\to\tau$ (i.e. near the Big Rip) the dependence of black
hole mass on $t$ becomes linear, $M\simeq\dot M_0\,(\tau-t)$.
While $t$ approaches to $\tau$ the rate of black hole mass
decrease does not depend on both an initial black hole mass and
the density of the phantom energy: $\dot M\simeq-\dot M_0$. In
other words masses of all black holes in the universe tend to be
equal near the Big Rip. This means that the phantom energy
accretion prevails over the Hawking radiation until the mass of
black hole is the Planck mass \footnote{Formally all black holes
in the universe evaporate completely at Planck time before the Big
Rip due to Hawking radiation}.

In remaining let us confront our results with the calculations of
(not phantom) scalar field accretion onto the black hole
\cite{Jac,BeaMag,FroKof,Unruh}. The dark energy is usually
modelled by a scalar field $\phi$ with potential $V(\phi)$. The
perfect fluid approach is more rough because for given 'perfect
fluid variables' $\rho$ and $p$ one can not restore the 'scalar
field variables' $\phi$ and $\nabla\phi$. In spite of the pointed
difference between a scalar field and a perfect fluid we show
below that our results are in a very good agreement with the
corresponding calculations of a scalar field accretion onto the
black hole.

The Lagrangian of a scalar field is $L=K-V$, where $K$ is a
kinetic term of a scalar field $\phi$ and $V$ is a potential. For
the standard choice of a kinetic term $K=\phi_{;\mu}\phi^{;\mu}/2$
the energy flux is $T_{0r}=\phi_{,t}\phi_{,r}$. Jacobson
\cite{Jac} found the scalar field solution in Schwarzschild metric
for the case of zero potential $V=0$:
$\phi=\dot\phi_\infty[t+2M\ln(1-2M/r)]$, where $\phi_\infty$ is
the value of the scalar field at the infinity. In \cite{FroKof} it
was shown that this solution remains valid also for a rather
general form of runaway potential $V(\phi)$.  For this solution we
have $T_0^{\;r}=-(2M)^2\dot\phi^2_\infty/r^2$ and correspondingly
$\dot M=4\pi(2M)^2{\dot{\phi}}^2_\infty$.

The energy-momentum tensor constructed from Jacobson solution
completely coincides with one for perfect fluid in the case of
ultra-hard equation of state $p=\rho$ under the replacement
$p_{\infty}\to\dot{\phi}_\infty^2/2$,
$\rho_{\infty}\to\dot{\phi}_\infty^2/2$ \cite{bde}. It is not
surprising because the theory of a scalar field with zero
potential $V(\phi)$ is identical to perfect fluid consideration
\cite{Luk80}. In a view of this coincidence it is easily to see
the agreement of our result (\ref{evol}) for $\dot M$ in the case
of $p=\rho$ and the corresponding result of \cite{Jac,FroKof}.

To describe the phantom energy the Lagrangian of a scalar field
must have a negative kinetic term \cite{Caldw}, for example,
$K=-\phi_{;\mu}\phi^{;\mu}/2$ (for the more general case of the
negative kinetic term see \cite{Gonz}). In this case the phantom
energy flux onto black hole has the opposite sign,
$T_{0r}=-\phi_{,t}\phi_{,r}$, where $\phi$ is the solution of the
same Klein-Gordon equation as in the case of standard scalar
field, however with the replacement $V\to-V$. For zero potential
this solution coincides with that obtained by Jacobson \cite{Jac}
for a scalar field with the positive kinetic term. Lagrangian with
negative kinetic term and $V(\phi)=0$ does not describe, however,
the phantom energy. At the same time, the solution for scalar
field with potential $V(\phi)=0$ is the same as with a positive
constant potential $V_0=const$, which can be chosen so that
$\rho=-\dot\phi^2/2+V_0>0$. In this case the scalar field
represents the required accreting phantom energy $\rho>0$ and
$p<-\rho$ and provides the decrease of black hole mass with the
rate $\dot M= -4\pi(2M)^2{\dot{\phi}}^2_\infty$.

The simple example of phantom cosmology (without a Big Rip) is
realized for a scalar field with the potential $V=m^2\phi^2/2$,
where $m\sim10^{-33}$~eV \cite{SamTop}. After short transition
phase this cosmological model tends to the asymptotic state with
$H\simeq m\phi/3^{1/2}$ and $\dot\phi\simeq2m/3^{1/2}$.  In the
Klein-Gordon equation the $m^2$ term (with the mentioned
replacement $V\to-V$) is comparable to other terms only at the
cosmological horizon distance. This means that the Jacobson
solution is valid for this case also. Calculating the
corresponding energy flux one can easily obtain $\dot
M=-4\pi(2M)^2\dot\phi^2_{\infty} =-64M^2m^2/3$. For
$M_0=M_{\odot}$ and $m=10^{-33}$~eV the effective time of black
hole mass decrease is $\tau=(3/64)M^{-1}m^{-2}\sim 10^{32}$~yr.

The possible physical interpretation of a black hole mass
diminishing is that accreting particles of phantom scalar field
have a total negative energy \cite{CliJeoMoo}. The similar
negative energy particles are created in the Hawking radiation
process and participate the Penrose black hole rotation energy
extraction mechanism. Formally saying the black hole mass
decreasing in the process of the phantom energy accretion is due
to the violation of the energy domination condition in phantom
energy. It should be noted that the existence of the horizon is
not crucial for the decrease of the black hole mass due to phantom
energy accretion. The one-way-membrane property of the horizon
merely makes this accretion inevitable and irreversible. Any
object or device would diminish its mass if it is capable to
capture phantom energy.

\begin{acknowledgments}

We acknowledge V.~Beskin, V.~Lukash, A. Doroshkevich and
V.~Mukhanov for helpful discussions and critical remarks. This
work was supported in part by the Russian Foundation for Basic
Research grants 02-02-16762-a and 03-02-16436-a.

\end{acknowledgments}

\end{document}